# Uncoupling Which-Way Information from Interference: A Novel Interference Experiment using a Super-Focused Laser Beam

Johan Wulleman

**Abstract**

The generally accepted view in quantum theory is that information about which way the quantum system traveled and interference visibility are complementary. In all which-way experiments, however, an intervention takes place in the interference process in order to determine which way the quantum system took. This intervention can imply the tagging of a which-way marker to a quantum system or, for instance, blocking off one of the paths in a Mach-Zehnder interferometer so that one indirectly knows that the quantum system took the other (open) path. It is, however, this intervention that destroys the interference. In this paper a novel two-slit which-way interference experiment will be discussed and proposed for implementation that provides maximum which-way information without intervening in the interference process so that simultaneously maximum interference visibility remains preserved. This, in fact, implies an uncoupling of which-way information from interference and consequently also entails violating the duality relation $P^2+V^2 \leq 1$. Basically, the purpose of the proposed experiment and of this paper is to scrutinize this duality relation. The experiment makes use of a super-focused laser beam that is launched into only one of the two slits of the two-slit interference experiment.

# 1 Introduction

One of the first epoch-making debates about which way a quantum system travels in a two-slit interference which-way thought experiment took place between Albert Einstein and Niels Bohr as early as 1927 during the fifth Solvay Congress in Brussels [1]. Later, Richard Feynman [2] proposed a two-slit which-way electron interference experiment with a light source behind the two-slit plate. Then, Scully *et al*. [3] proposed a two-slit which-way thought experiment with atoms which resulted in a debate [4–7] about what caused the disappearance of the interference: entanglement or momentum kicks. More recently Dürr *et al*. [8] performed an atom which-way interferometer experiment and concluded that in their experiment entanglement led to the destruction of the interference pattern. Of course, which-way experiments have also been performed earlier using photons [9–11] and with similar results as with atoms [12] and even neutrons [13–15]. Hence, the general believe is that information about which way the system traveled and interference visibility are complementary, giving rise to the duality relations $P^2+V^2 \leq 1$, $K^2+V^2 \leq 1$ and $D^2+V^2 \leq 1$ [16–22].

Noteworthy to mention is that the common observation in all which-way interference experiments is that there is an intervention in the interference process itself thereby destroying the interference. This intervention may imply a direct (which-way) interaction with the interfering quantum system, for instance by adding a which-way marker to the quantum system, which in the case of photons can be a polarization direction. On the other hand, the intervention in the interference process may, for instance, also result from a blockage of one of the paths in a Mach-Zehnder type Interferometer (MZI); this way one indirectly knows, without the need for any direct which-way interaction with the quantum system in the interferometer, that upon detection at the output the quantum system took the open path. Nonetheless, also here the interference is destroyed due to the intervention in the interference process. The same reasoning also holds for a strongly asymmetric MZI: one knows what path the quantum system took but all interference at the output is gone due to the fact that 'by construction' of the experimental set-up the strongly asymmetric MZI suppressed one of the two paths thereby intervening in the interference process. So, if one wants to acquire information about which way the quantum system took, one intervenes somehow in the interference process thereby destroying the interference. Hence, which-way information and interference are complementary.

In section 2, however, I will discuss and propose for implementation a novel which-way two-slit interference (thought) experiment that provides absolute certainty about which way the interfering quantum systems (photons) traveled *without* intervening in the interference process. The purpose of the proposed experiment is to acquire maximum which-way information and simultaneously maximum interference visibility. The experiment makes use of a super-focused laser beam that can be focused to a tiny sub-wavelength spot size [23–32] and which is then launched into only one of the two slits of the two-slit plate. In section 3 the consequences of the experiment's expected outcome are discussed. It is argued the proposed experiment will upon experimental realization give rise to simultaneous maximum which-way information and maximum two-slit interference visibility, consequently violating the duality relation $P^2+V^2 \leq 1$. In the last section, the conclusions, a short summary is given about what impact the violation of the duality relation may have in the debate on the foundations of quantum physics. Just as Brida *et al.* [33] have shown a serious flaw in the de Broglie-Bohm theory, this proposed experiment may upon experimental confirmation indicate that some aspects of the standard quantum theory aren't without any shortcomings either.

## 2 Novel two-slit which-way experiment

### 2.1 Focusing a laser beam to a tiny spot

The actually proposed two-slit experiment relies on super-focusing where a laser beam can be focused to a tiny focal spot with a diameter of just a few wavelengths or less. Theoretical work [23–25] already predicted one could get around the diffraction limit and recent experiments in this field [26–28] even demonstrated a three-dimensional (3D) sub-wavelength focal spot size (usually determined by the $1/e^2$ intensity point from the maximum; see any text book on laser optics) of well below one wavelength. For our proposed two-slit experiment, the disadvantage of 3D sub-wavelength focusing is the too large focusing angle $\theta$ in the focal spot, which implies a spread in the direction of the photon trajectories over a total range of $2\theta$. Now, this spread in the direction of the photon trajectories can wash out completely any two-slit interference pattern when the focused beam is directed onto the two-slit plate. Very fortunately there is no need for 3D focusing in our proposed two-slit experiment, only focusing along the transverse direction perpendicular to the slits, i.e. focusing along the *x*-direction in Fig. 1, is required.

Furthermore, the spot size in the *x*-direction does definitely not need to be of a sub-wavelength size − in fact, along the *x*-direction the spot needs only to have a width in the order of just a few (or several) wavelengths, i.e. a total spot width in the order of or less than *d*. As there are no requirements set for the spot size along the axial (*z*-)direction nor along the transverse *y*-direction, one can at the expense of the spot size and resolution along the axial and *y*-direction [25] focus the laser beam in the *x*-direction to the order of a few wavelengths while maintaining an extremely small focusing angle $\theta$ (see Fig. 1). The fact that the focusing angle $\theta$ must remain as small as possible is an important issue in the proposed experiment as it ensures a highly collimated beam where the photon trajectories are mutually quasi parallel: a highly collimated beam with small $\theta$ is a prerequisite for observing any two-slit interference pattern. To give the reader an idea about the value, $\theta$ should be at least an order of magnitude smaller than $\phi$ where the latter is the angle that determines the position of the first minimum in the two-slit interference pattern with $\sin\phi \approx \lambda/2d$ [see Fig. 2(b)]. For instance, with $\lambda = 632.8$ *n*m the wavelength of the (He-Ne) laser and $d = 12.6$ $\mu$m, we find that $\phi \approx 25$ *m*rad. Hence, $\theta$ should then be in the order of, let us say, 1 *m*rad. If that condition is not fulfilled and $\theta$ is in the order of or larger than $\phi$, the two-slit interference pattern will be totally blurred. A remark of practical interest is to use an in the *y*-direction linear polarized laser beam as focusing perpendicular to the polarization direction is slightly better than along the direction of polarization [27]. As $d = 12.6$ $\mu$m, a spot width of, say, $10$ $\mu$m ($<d$) would be small enough to perform the experiment.

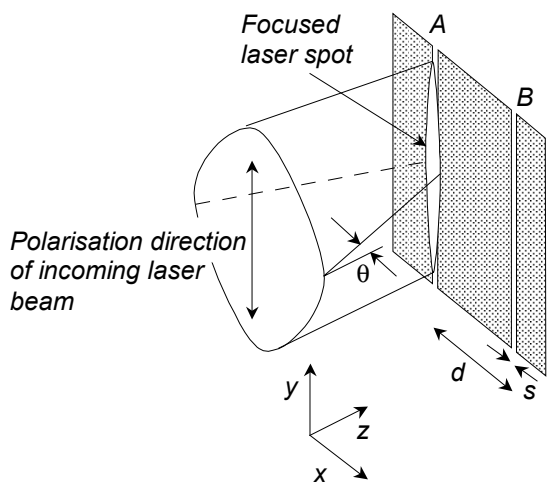

Figure 1: Laser beam one-dimensionally focused in the *x*-direction to a spot width of just one or a few wavelengths. Note the small focusing angle $\theta$ and the vertical polarization of the incoming laser beam.

The combination of a small spot width and a small $\theta$ is maybe very evocative to some as it looks like we are dealing with a 'diffraction-free' beam, but in fact diffraction-free (or quasi diffraction-free) beams exist almost two decades now and is nothing new. Those diffraction-free beams have a transverse intensity profile that obeys a squared zeroth-order Bessel function $J_0^2$. One way to implement a Bessel beam is to expand a laser beam (which usually has a quasi Gaussian transverse intensity profile) by using a beam expander, then direct the expanded beam to a ring (annular) slit that is placed in the back focal plane of a lens; the lens itself then focuses the light to a Bessel beam on the front side of the lens. Note that a Bessel beam can be considered as a Fourier transform of a ring – here the lens acts as a Fourier transformer. The intensity profile of the Bessel beam is composed of a narrow high intensity central core surrounded by concentric rings with significantly lower intensity than the core; see for instance the work of McQueen *et al.* [29] and McGloin *et al.* [30] for representations of Bessel beams. The main properties of a Bessel beam is that the core can be focused to a small spot size in the order of and even less than 10 $\mu$m and remain focused over long distances. It was Durning *et al.* [24] who worked out the theoretical background and experimentally demonstrated that a Bessel beam remained focused at a core diameter of 70 $\mu$m over a distance of 70 cm (0.7 m) without appreciable spreading. For comparison, the Rayleigh range of a Gaussian beam, given by [29] $Z_R = \pi w_o^2/\lambda$ with $w_o = 70\,\mu$m the beam waist and $\lambda = 633\,n$m, is 2.4 cm. McQueen *et al.* demonstrated that with ordinary laboratory equipment one could keep a Bessel beam focused to a spot diameter of 20 $\mu$m over a distance of 50 cm. Instead of using a ring slit and lens, one can use an axicon, which is also known as conical lens or rotationally symmetric prism. Axicons are widely used and one of the advantages is that the rapid intensity oscillations along the on-axis propagation of the Bessel beam that are seen in the work of Durning *et al.* are removed giving a smoother intensity variation along the on-axis [30]. This improves the collimation and so the mutual parallelism of the photon trajectories in the core of the Bessel beam and that is exactly what we need: parallel photon trajectories in the core. The strong collimation of the core of a Bessel beam with a diameter of 4 $\mu$m and collimated over a distance of 3 *m*m is demonstrated experimentally and by numerical simulations in the work of Dholakia *et al.* [31]. Even when a photon trajectory would run skew by 4 $\mu$m over a distance of 3 *m*m, still this corresponds with a negligible angle of 4 $\mu$m/3 *m*m ≈ 1.3 *m*rad, which is an order of magnitude smaller than the critical angle of 25 *m*rad. So the idea is then to adjust the focusing so that the core of the Bessel beam strikes slit *A* and the first dark ring strikes slit *B* in Fig. 1. About Bessel beams there is, however, still a noteworthy remark to make. It is shown [32] that these beams have planar phase fronts with a π-phase shift from one Bessel ring to the next ring and so also between the core and the first Bessel ring. This implies a π/2-phase shift between the core and the first dark ring and, henceforth, implies a π/2-phase shift between the wavefronts at slit *A* and slit *B*. However, as the phase difference between slit *A* and slit *B* is constant in time, this phase difference is not a problem for observing any two-slit interference pattern at a screen. The only consequence of this phase difference is that the two-slit interference pattern at the screen is shifted along the *x*-axis as if the incident Bessel beam has been slightly tilted with respect to the normal (*z*-axis) of the two-slit plate under a very small angle approximated as $(\lambda/4)/d \approx (623\,n\text{m}/4)/12.6\,\mu\text{m} \approx 13\,m$rad within the *x-z* plane. Another important issue is that the two-slit plate must be made of a 100% absorbing material, i.e. the reflectivity *R* should be approximately zero. If that is not the case then part of the incident beam will reflect back from the two-slit plate and interfere with the incoming beam thereby scattering and disturbing the beam core just in front of the slits. That would be especially the case for a Bessel beam as its *k*-vector propagates on a cone. A good review about the foundations of Bessel beams and their applications is given in McGloin *et al*. [30].

Summarizing for the issue of focusing a laser beam, from Refs. 23–32 it is clear that the basic technology is definitely available to focus a laser beam to a spot width of, say, $10\,\mu$m ($<d$) that is small enough to perform the experiment while simultaneously retaining a focusing angle significantly smaller than the critical angle of 25 *m*rad.

**2.2 Novel particle-wave model**

Next to discuss is the novel particle-wave composite model in Fig. 2(a) that I envisage for a quantum system. In what follows, the foundations and fundamentals of this novel particle-wave model will be laid down and qualitatively discussed. Now, this particle-(de Broglie)wave model in Fig. 2(a) is deduced from in open literature published interference experiments using photons as well as massive particles. It is not my intention in this paper to develop in full detail a new particle-de Broglie wave model. Instead, I will qualitatively describe the novel particle-de Broglie wave model to a level of enough detail so that the model can be used to quantitatively describe the interference pattern of Fig. 2(b). So, to characterize and to get a clearer picture about my particle-de Broglie wave model we will have a look at interference experiments using neutrons, atoms and photons. Note that the intensity of the incident particle beam is low enough so that there is only one single particle at the time in the interferometer. That way we are sure the interference is clearly self-interference. In my particle-de Broglie wave model in Fig. 2(a) the corpuscle particle-part (black dots) is accompanied by a physically real de Broglie wave (straight plane lines).

From two-slit neutron [34] or atom [35] interference experiments we learn that the particle-part (and its joint physical quantities like mass, spin momentum and charge) of the individual neutron or atom quantum system passes only one of the two slits, i.e. the particle-part is not split during passage through the slits. Just imagine the consequences if that would not be the case and the particle-part of a neutron or the nucleus of an atom is indeed split during passage through the two-slit plate?! Basically this would mean that the interference experiment would be an open fission reactor – of course, the particle-part is not split over the two slits and neither is it split over the two arms in the case of an atom [12] or neutron [17] Mach-Zehnder Interferometer (MZI). Also in the case of a MZI using photons the photon particle-part is definitely not split over the two arms and follows only one path along one of the arms, the other arm does not conduct any particle-part. The latter can be inferred from the fact that if two detectors are put behind the input beam splitter, one in each arm, then (a) only one of the detectors will fire after a photon entered the input beam splitter and (b) the measured photon energy in the detector that fired is the same as the energy of the photon just before it entered the input beam splitter.

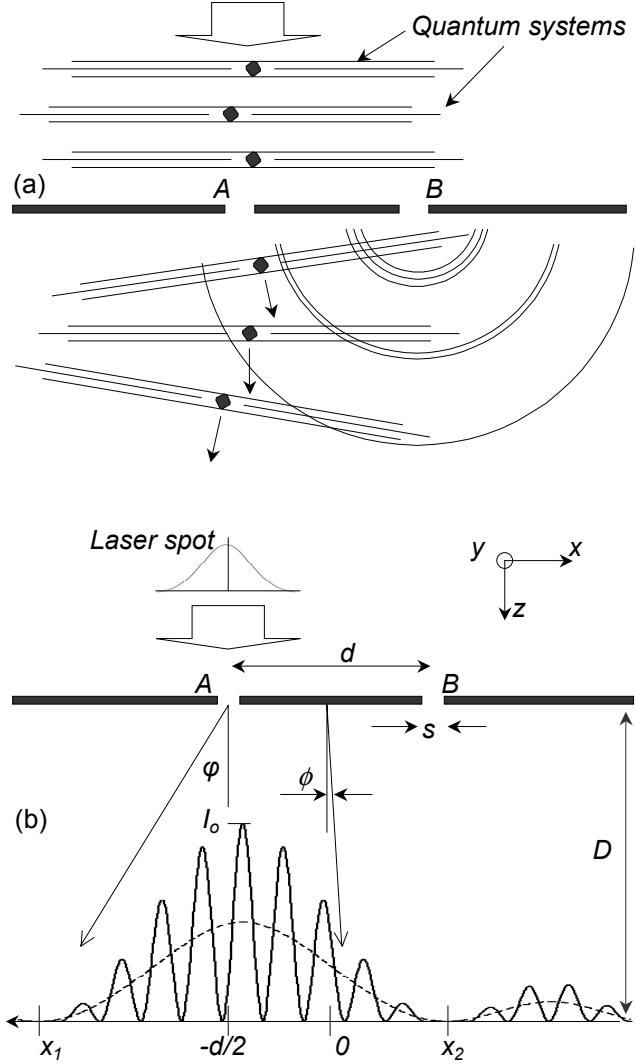

Figure 2: The final two-slit experiment with a laser focused only onto slit *A*: (a) The quantum systems (photons) are represented by their particle-part (black dot) and their accompanying de Broglie wave (horizontal lines). In Fig. 1 as well as this figure the laser beam spot size is determined by the photon's particle-part while the de Broglie waves stretch out far wider than the laser spot (beam) width; (b) The final two-slit interference pattern (solid line) is the result of the $\cos^2()$ term [Eq. (10)] being modulated onto the from slit *A* originating single-slit diffraction pattern (dotted curve with its maximum opposite slit *A*). The angles $\phi$ and $\varphi$ can be approximately determined by the relations $\sin\phi = \lambda/2d$ and $\sin\varphi = \lambda/s$. Note that all curves in Figs. 2 and 3 are not true in scale.

Hence, for a photon MZI the photon's energy, and so the photon itself (particle-part) with all its physical quantities like the electric (E) and magnetic (B) field, is definitely not split over the two arms and goes entirely along only one of the two arms. The same holds for two-slit interference using photons: i.e. the photon particle-part, with all the photon's energy and physical quantities, passes only one of the two slits. The latter finding can be directly deduced from the interference pattern at the screen. If for some reason the photon particle-part would be split in two so that the photon's energy would be equally distributed over the two slits, then this would imply that two photons with half the original energy each pass a different slit. As a result of the energy divided by half, both new photons would exhibit a wavelength twice the original wavelength so that the interference pattern at the screen would definitely not correspond with the wavelength of the original photon. As we observe at the screen an interference pattern that corresponds with the original wavelength before the photon entered the two-slit interferometer, one is left with the conclusion that all the photon's energy, and so its particle-part with all its physical quantities, must have passed only one of the two slits during passage. Henceforth, the photon particle-part in a two-slit interference experiment is definitely not split over the two slits and passes only one of the two slits.

The interference pattern at the output detector behind the output beam splitter of the MZI can only be explained in terms of waves: two different (de Broglie) waves each going along the different arms reach simultaneously the output beam splitter. Then, knowing that in a MZI the corpuscle particle-part is not split over the two arms, one is left with the unavoidable conclusion that from the input beam splitter onwards the particle-part follows a path along one arm, and that at the input beam splitter the de Broglie wave of the incoming quantum system is separated in two: i.e. (a) one part of de Broglie wave joints the particle-part and (b) the other part of the de Broglie wave follows the other path along the other arm as an empty de Broglie wave. Both the particle-part (with its joint de Broglie wave) and the empty de Broglie wave then simultaneously meet at the output beam splitter where they locally interfere. So, at the input beam splitter the de Broglie wave is separated in two and one part keeps on joining the particle-part while the other part follows the other path (arm) as an empty de Broglie wave. The same reasoning holds for two-slit interference. The two-slit interference pattern can only be explained by two coherent (de Broglie) waves simultaneously emanated from the two slits. Knowing that the particle-part of the quantum system passes only one of the two slits while simultaneously two coherent de Broglie waves, one for each slit, pass both slits, then the conclusion is that one of the de Broglie waves must pass one of the two slits together with the particle-part and that the other de Broglie wave through the other slit must be an empty de Broglie wave. This reasoning holds for all quantum systems, massive ones as well as for photons.

I would like to stress the important fact that a (empty) de Broglie wave is free of any physical quantity. As mentioned earlier, the particle-part (mass, spin momentum and charge) of the individual quantum system passes only one of the two slits or arms in the case of a MZI, i.e. the particle-part is not split during passage through one of the two slits nor is it split over the two arms. Atoms, for instance, are clearly not split over the two slits or arms during passage. That is to say, all the atom's physical quantities like its mass, charge and spin pass only one of the two slits or arms so that by construction the empty de Broglie wave through the other slit or arm is free of any physical quantity and therefore literally empty. And, as there is no reason why we should treat other quantum systems differently, the finding that a (empty) de Broglie wave is free of any physical quantity does not only hold for atoms but for any type of quantum system, including massive quantum systems and photons. For a photon, this implies that the (empty) de Broglie wave does not carry any electric (E) nor any magnetic (B) quantity; i.e. the E and B quantity is carried solely by the particle-part of the photon. The fact that empty de Broglie waves do not carry any E nor any B quantity clearly manifests itself in the fact that empty de Broglie waves do not induce coherence [36].

As the literally empty de Broglie wave clearly plays a crucial role in the formation of two-slit and MZI interference, one has to conclude that this interference is a pure de Broglie wave phenomenon where in the case of photons the E and B quantity does not play any role whatsoever in the creation of the interference. That interference using photons is a pure de Broglie wave phenomenon has been clearly demonstrated by D'Angelo *et al*. [37] and Edamatsu *et al*. [38]. One might be tempted to believe that on the level of the physical reality single photon two-slit or MZI interference is due, somehow, to the presence of the E and B quantity (or field), but this is definitely not the case. Just as with neutrons and atoms, single photon two-slit and MZI interference is also a pure de Broglie wave phenomenon.

What else do we learn from the MZI? Well, if in one of the arms a blocking device is put so that the quantum systems in that arm are completely stopped, the total number of particles that reach the output beam splitter will be reduced by half. But most importantly also the interference is gone for those particles reaching the output beam splitter. This can only be explained as follows: as the particles that reached the output beam splitter came via the non-blocked arm, then the empty de Broglie waves in the other arm must have been stopped as well by the blocking device. This brings us to the conclusion that an empty de Broglie wave in some respect behaves as the normal quantum system where it was originally deduced from; i.e. the stopping cross section of the original quantum system and of its empty de Broglie wave are the same. In other words, if the original quantum system can be stopped by an absorbing device, then so will its empty de Broglie wave be stopped the same way. This finding will be needed in the further discussion of the two-slit interference.

So, for any quantum system's (neutrons, atoms, photons, electrons or even molecules) particle-part that passes through slit *A* in Fig. 2(a), that part of its original de Broglie wave that collides with the blocking material of the two-slit plate will be absorbed and therefore stopped. The part of the de Broglie wave at slit *B*, will pass slit *B* over a width *s* and continue to propagate and to expand behind slit *B* as an empty de Broglie wave. Simultaneously, the particle-part passes slit *A* joint by that part of the de Broglie wave that remains connected to the photon's particle-part and which I call the non-empty de Broglie wave. At this point the original quantum system has been separated into the particle-part joint by the non-empty de Broglie wave and into an effectively empty de Broglie wave. From some distance onwards behind the two-slit plate the empty de Broglie wave from slit *B* will then meet the particle-part and interact with it. As the particle-part continues to move away from the slit plate, the particle-part is interacting not only with its own non-empty de Broglie wave but also with the empty de Broglie wave. This interaction process then congregates the traveling particle-parts into particular directions forming a two-slit interference pattern at the screen. This interaction process and its outcome are valid for any two-slit interference, no matter if all individual particle-parts always pass slit *A* or pass sometimes slit *A* and then slit *B* as in any classic two-slit self-interference experiment.

Now, an intriguing question is how far does the de Broglie wave's influence stretch out beyond the particle-part. Well, also this can be deduced from ordinary interference experiments. In any two-slit self-interference photon experiment (see Fig. 3) with realistic figures (e.g. $\lambda = 633\,nm$ and $d = 12.6\,\mu m$), a two-slit interference pattern with nearly 100% visibility ($V = 1$; see Eq. (17) below) will build up at the target screen, even if the distance between the two slits is increased to the order of 50 times (or higher) the photons' wavelength. Knowing that the particle-part passes only one of the two slits, then from this we infer that the noticeable influence of the de Broglie wave stretches out in space at least 50 wavelengths away from the particle-part of the photon. Hence, in my particle-de Broglie wave model that I envisage in Fig. 2(a), the de Broglie wave stretches then out over a distance in transverse direction of at least several tens of times the wavelength away from the particle-part of the quantum system. Obviously there isn't any clear cut distance at which the amplitude of the de Broglie wave drops off sharply; by increasing the distance between the two slits one can clearly observe that the two-slit interference pattern fades away gradually. A practical consequence of this finding is that any interference phenomenon (single-slit, two-slit, edge) starts to fade away gradually from some distance onwards between the photon's particle-part and the relevant edge.

Another important conclusion from the fact that the interference pattern exhibits a nearly 100% visibility is that after passage of the respective slits *A* and *B* (with equal width *s*) the amplitudes of the non-empty and the empty de Broglie wave are approximately equal. If that would not be the case and the amplitude of the empty de Broglie wave goes to zero then the two-slit self-interference pattern would vanish. As the latter is not the case in any ordinary two-slit self-interference experiment and a two-slit pattern with a 100% visibility is obtained, the only plausible explanation is that the amplitude of the empty de Broglie wave must be approximately equal to the amplitude of the non-empty de Broglie wave upon passage through the slits. I will come back to this issue in the context of the factors $\alpha$ and $\beta$ in Eq. (2).

Summarizing, my particle-(de Broglie )wave composite model in Fig. 2.(a) for the *single* quantum system is composed of physically two real objects (i.e. they both objectively exist), namely: (a) an undividable corpuscle particle-part and (b) an accompanying de Broglie *wave* which stretches out several tens of wavelengths away from the particle-part. This de Broglie wave does not carry any physical quantities and part of this de Broglie wave can be split off resulting in an empty de Broglie wave which on its turn can then interact with the particle-part of the original quantum system so to congregate the directions of the latter into an interference pattern.

Without going into a detailed analysis, it should be noted that my particle-de Broglie wave model should not be confused with the fundamentally different "pilot-wave $\psi$" idea of de Broglie which is the basis of the de Broglie-Bohm theory [39] where, for instance, trajectories of individual quantum systems are calculated for the case of two-slit interference (see Fig. 5.5 in Ref. 39). In de Broglie's "pilot-wave $\psi$" model the mathematical wavefunction $\psi(x,y,z,t)$ is the solution of Schrödinger's wave equation and it is claimed that $\psi$ is physically real and guides the corpuscle particle-part. I definitely do not share that view. Interference is about self-interference on the level of the single quantum system and so it is, according to my opinion, obvious and logical to use mathematical descriptions that are directly related to the single quantum system, i.e. descriptions of the corpuscle particle-part and the accompanying de Broglie wave which are both physical real objects on the level of the single quantum system. On the other hand, $\psi(x,y,z,t)$ in the "pilot-wave $\psi$" model has no direct real physical relation with the single quantum system as such but only with an *ensemble* of quantum systems. So it is in my view very illogical to use $\psi$ as guidance for the particle-part of a single individual quantum system. Whether $\psi$ is to be considered ontological (i.e., the actual situations) or epistemological (what we can know) is at first instance of no relevance; the first and real problem with $\psi$ is that it is related to an *ensemble* of quantum systems and not directly to the single individual quantum system itself. Putting it in slightly different words, $\psi$ does not directly represent the physical reality of the individual quantum system. So in my model $\psi$ does definitely not at all guide the particle-part; instead, it is the interaction between, on one hand, the individual particle-part and, on the other hand, the individual and physically real empty and non-empty de Broglie waves that congregates the particle trajectories into particular directions towards an interference pattern at the screen. In fact, in my particle-de Broglie wave model there are three issues that play a role: first, the corpuscle particle-part; second, the physically real de Broglie wave and, third, a wavefunction $\psi(x,y,z,t)$ which is a pure and only mathematical instrument that represents a probability amplitude for finding the corpuscle particle at a particular time $t$ in some region $(x,y,z)$ between the two-slit plate and the screen.

The main issue here for the reader is not to confuse between my model and de Broglie's "pilot-wave" idea, despite their resemblance. Let me emphasize again that in this paper it is not my intention to develop in extensive detail a new de Broglie wave model. At this point, the description of my particle-de Broglie wave model that I envisage is sufficiently detailed enough to describe quantitatively the interference pattern in Fig. 2(b), as the latter is the essence of this paper.

**2.3 Revised mathematical analysis**

Let us now make an attempt to mathematically describe the two-slit self-interference process on the level of an individual photon (or any other quantum system for that matter). Upon the photon's particle-part passage through slit $A$, the particle-part will at first experience a single-slit diffraction because right after its passage through slit $A$ the empty de Broglie wave from slit $B$ cannot reach instantaneously the particle-part at slit $A$ as this would require an infinitely high velocity from the empty de Broglie wave, and the latter can clearly not be the case. Note that I assume that the velocity of the empty de Broglie wave is the same as the velocity of the particle-part at the moment of separation, i.e. upon passage through the slits. So, the particle-part upon passage through slit $A$ undergoes first a single-slit diffraction, as if slit $B$ is non-existing, from which the to the screen extrapolated wavefunction is represented as $\Psi_{single,A}$. Only further away from the two-slit plate the particle-part experiences, on top of the single-slit diffraction, a two-slit interference for which the associated wavefunction at the screen can be approximated as $\Psi_A + \Psi_B$. Herein $\Psi_A$ and $\Psi_B$ represent the wavefunctions at the target screen from respectively the slits $A$ and $B$ with the approximation that slits $A$ and $B$ are each replaced by a single point source of empty de Broglie waves that are put in the center of their corresponding slits and where both sources are mutually in-phase.

The next (and most) important step is then how to "bring together" at the screen the functions $\Psi_{single,A}$ and $\Psi_A + \Psi_B$? For this I suggest – and this is a crucial step – that on the level of physical reality the two-slit interference is modulated onto the single-slit diffraction as the particle-part first experiences a single-slit diffraction and only further away from the two-slit plate experiences then the two-slit interference. Mathematically this is translated by simply multiplying the terms $\Psi_{single,A}$ and $\Psi_A + \Psi_B$ so that the total wavefunction, $\Psi_{t,A}$, in Fraunhofer regime at the target screen can be approximated as

$$\Psi_{t,A} = N\Psi_{single,A}(\Psi_A + \Psi_B) \qquad (1)$$

where $\Psi_{single,A}$ is the single-slit wavefunction from slit $A$ as if slit $B$ is non-existing. $N$ is a normalization factor who's value is determined by the condition $\int \Psi^*_{t,A}\Psi_{t,A}dx = 1$ with $dx$ going along the screen. So the term $(\Psi_A + \Psi_B)$ gives later on rise to the two-slit interference being modulated onto the single-slit diffraction pattern from slit *A*. Hence, $\Psi_A$ and $\Psi_B$ are given in the general form by

$$\Psi_A = \alpha\, e^{ikr_A}, \ \Psi_B = \beta\, e^{ikr_B} \qquad (2)$$

with $\alpha$ and $\beta$ real-valued normalization constants which can be transferred into *N* in our case because $\alpha \approx \beta$; $r_A = |\mathbf{r}_A|$ and $r_B = |\mathbf{r}_B|$ are the absolute distances between a particular point at the screen and the center of their corresponding slit. As mentioned above, any real two-slit self-interference pattern with a nearly 100% visibility, that is *d* is small enough, entails that $\alpha$ and $\beta$ must be approximately equal, $\alpha \approx \beta$. If not, and $\beta$ tends to go to zero, then in any two-slit interference experiment the visibility in the two-slit pattern on the screen would also tend to go to zero [see Eq. (9)]. From experimental observations the latter is clearly not the case in two-slit interference and henceforth $\alpha$ and $\beta$ must be approximately equal. Furthermore, whether we deal with an experimental set-up where the photons' particle-part enter sometimes slit *A* or slit *B*, or the particle-part enters always slit *A*, as in Fig. 1, is of no relevance as in both situations the interference is about self-interference. In other words, $\alpha \approx \beta$ definitely holds for the proposed set-up in Figs. 1, 2 and 3.

The correctness of Eq. (1) can be easily verified. Take now the situation where the width of single slit *A* tends to zero while still letting through the incident photons. This is equivalent with single slit *A* being replaced by a photon-ejecting de Broglie wave point-source with unit strength (amplitude) which is in-phase with the de Broglie wave from slit *B*. This situation eliminates any single-slit pattern at the screen; that is to say, the first order zero intensity points at $x_1$ and $x_2$ of the single-slit curve [dotted curve in Fig. 2(b)] go to respectively – and + infinity and the intensity at the region of interest on the screen is approximately constant and hence uniformly distributed along *x*. In this case $\Psi_{single,A}$ at the screen becomes

$$\Psi_{single,A} = \zeta\, e^{ikr_A} \qquad (3)$$

where $\zeta$ is a normalization constant that can be transferred into *N*. From Eq. (1) the probability density for finding photons at a particular *x* co-ordinate along the screen is then given as

$$\Psi_{t,A}^* \Psi_{t,A} = N^2 e^{-ikr_A} e^{ikr_A} (\Psi_A^* + \Psi_B^*)(\Psi_A + \Psi_B) \qquad (4)$$

which then finally results in the correct expression for the two-slit interference being

$$\Psi_{t,A}^* \Psi_{t,A} = N^2 (\Psi_A^* + \Psi_B^*)(\Psi_A + \Psi_B). \qquad (5)$$

Also the other limit case, i.e. single-slit diffraction, can be verified. The total closure of slit B and the removal of its point source is represented by taking the normalization constant $\beta$ in Eq. (2) equal to zero; $\beta = 0$. From Eq. (1) the probability in this case for finding photons at a particular $x$ co-ordinate along the screen is then given as

$$\Psi^*_{t,A}\Psi_{t,A} = N^2 \Psi^*_{single,A}\Psi_{single,A}\Psi^*_A\Psi_A \tag{6}$$

and with $\Psi_A \propto \exp(ikr_A)$ this becomes

$$\Psi^*_{t,A}\Psi_{t,A} = N^2 \Psi^*_{single,A}\Psi_{single,A} \tag{7}$$

which represents simply the probability distribution on the screen for single-slit diffraction coming from slit $A$. Hence, the wavefunction in Eq. (1) describes the limit cases of single-slit diffraction and pure two-slit interference. Note that Eq. (7) represents what slit the quantum system took, and that is slit $A$, so that the single-slit diffraction term gives us in fact indirectly absolute knowledge which way (slit) the quantum system's (photon's) particle-part took.

For the two-slit situation of Fig. 2(b) the probability density from Eq. (1) is given as

$$\Psi^*_{t,A}\Psi_{t,A} = N\Psi^*_{single,A}\Psi_{single,A}(\Psi^*_A + \Psi^*_B)(\Psi_A + \Psi_B). \tag{8}$$

Using Eq. (2) the two-slit interference term results in

$$(\Psi^*_A + \Psi^*_B)(\Psi_A + \Psi_B) = \alpha^2 + \beta^2 + 2\alpha\beta\cos(k(r_A - r_B)) \tag{9}$$

and taking into account that $D \gg d$, that both slits have the same width $s$ and that $\varphi$ [see Fig. 2(b)] is relatively small so that the amplitude of both waves from slit $A$ and $B$ at the screen are approximately equal ($\alpha \approx \beta$), then the right hand side of this expression finally becomes [40]

$$1 + \cos(k(r_A - r_B)) = 2\cos^2\left(\frac{\pi x d}{\lambda D}\right) \tag{10}$$

where the constants $\alpha \approx \beta$ have been transferred into $N$ and use was made of the relation $1 + \cos(\omega) = 2\cos^2(\omega/2)$. On the other hand, the term representing the single-slit interference in Fraunhofer regime can be approximated as [40]

$$\Psi^*_{single,A}\Psi_{single,A} \propto \left[\frac{\sin(\pi s(x + d/2)/\lambda D)}{\pi s(x + d/2)/\lambda D}\right]^2 \tag{11}$$

with $s$ the single slit width and the center of slit $A$ at $x=-d/2$. Eq. (11) is represented as the dotted curve in Fig. 2(b) and as curve $a$ in Fig. 3(a). Eq. (8), which is also the intensity, then finally results in

$$\Psi^{*}_{t,A}\Psi_{t,A} = I_A(x) = I_0 \left[ \frac{\sin(\pi s(x+d/2)/\lambda D)}{\pi s(x+d/2)/\lambda D} \right]^2 \cos^2\left(\frac{\pi x d}{\lambda D}\right) \quad (12)$$

where the normalization constant $N^2$ is absorbed by $I_0$, which is the maximum intensity at $x \approx -d/2$ on the screen and where $I_0 \approx 2I_{0,single}$ with $I_{0,single}$ the maximum intensity of the from slit $A$ coming dotted curve at $x=-d/2$ in Fig. 2(b). The two-slit interference intensity pattern $I_A(x)$ is represented by the solid-line curve in Fig. 2(b). An important remark maybe for Fig. 2(b) is that Heisenberg's single-slit uncertainty relation $s\Delta p \geq h$ remains upheld. The reason for this is the fact that the photon particle-parts in Fig. 2(b) can pass slit $A$ anywhere within the width $s$ – i.e. there is an uncertainty of $\Delta x = s$ about where exactly the photon particle-part passes slit $A$. That the uncertainty relation $s\Delta p \geq h$ remains upheld is of no surprise as most of the intensity in Fig. 2(b) is spread out but still confined mainly between $x_1$ and $x_2$. So, basically we have in Fig. 2(b) a single-slit interference intensity pattern with its center opposite slit $A$, and which is then modulated by a two-slit interference due to the empty de Broglie waves from slit $B$. Eqs. (8) and (12) both represent the probability density function for finding the particle-part, of an individual photon (or any other quantum system) that went through slit $A$, at a given $x$ co-ordinate along the screen. In case the laser beam would be focused onto slit $B$, instead of slit $A$, then we find the symmetric intensity pattern given as

$$I_B(x) = I_0 \left[ \frac{\sin(\pi s(x-d/2)/\lambda D)}{\pi s(x-d/2)/\lambda D} \right]^2 \cos^2\left(\frac{\pi x d}{\lambda D}\right) \quad (13)$$

with the center of slit $B$ at $x=+d/2$ and where the part $[\sin()/()]^2$ is represented by the in Fig. 3(a) dotted curve $b$ which is shifted over a distance $d$ with respect to curve $a$, i.e. $|x'_1 - x''_1| = d$. Noteworthy to mention is that the experiment of Fig. 2 can also be implemented with a focused atom beam instead of a laser beam. A proposal for directing a focused atom beam into a narrow slit, using a Scanning Tunneling Microscope, has been discussed in Ref. 41.

We have come to a point now where we can summarize and elucidate some aspects of the interpretation of $\Psi_{t,A}$ in Eq. (1). Knowing that the particle-part always passes slit $A$, it is clear that $\Psi_B$ in Eq. (1) is the wavefunction of a physically real empty de Broglie wave coming from slit $B$. The effect of the non-empty de Broglie wave joint by the particle-part when passing through slit $A$ can be split in two, $\Psi_A$ and $\Psi_{single,A}$, in which $\Psi_{single,A}$ represents the quantum system's particle-part joint by a non-empty de Broglie wave and where $\Psi_A$ represents an in-phase empty de Broglie wave just as $\Psi_B$ does. Now, the splitting into $\Psi_A$ and $\Psi_{single,A}$ may with some readers rise the question how to determine the maximum amplitude of the separate wavefunctions $\Psi_A$ and $\Psi_{single,A}$. This problem doesn't raise itself in Eq. (1) for the simple reason that $\Psi_A$ and $\Psi_{single,A}$ appear in a multiplicative form, i.e. $\Psi_A\Psi_{single,A}$, and so their amplitudes can be absorbed in the normalization constant $N$. If, on the other hand, $\Psi_A$ and $\Psi_{single,A}$ would have come up as a sum, i.e. $\Psi_A+\Psi_{single,A}$, then the question about how to determine their amplitudes would indeed cause a problem, but fortunately that is not the case in Eq. (1). $\Psi_A$ interferes with $\Psi_B$ which leads to the two-slit interference while $\Psi_{single,A}$ gives rise to the single-slit diffraction coming from slit $A$. Important to see is that there is first the single-slit diffraction carrying the particle-part of the quantum system when it passed slit $A$, and only later there is the two-slit interference being modulated onto the single-slit diffraction. So, appropriate factorization of $\Psi_A$ and $\Psi_{single,A}$ in Eqs. (1) and (8) leads to a factorization of single-slit diffraction and two-slit interference terms in Eqs. (12) and (13). Furthermore, as can be deduced from Eq. (7) (which represents the single-slit diffraction term) what slit the particle-part took thereby giving us absolute which-way(slit) information, Eq. (1) in fact entails indirectly the uncoupling (factorization) of the particle-part, which passes through slit $A$, from the wave part that passes through both slits where the latter gives rise to the two-slit interference.

Another noteworthy issue to elaborate is the fundamental difference in the interpretation between, on one hand, $\Psi_A$ and $\Psi_B$ and, on the other hand, $\Psi_{single,A}$. The generalized form of the wavefunction for a quantum system relevant for our two-slit experiment, is given as $\Psi(\mathbf{r})=|\Psi(\mathbf{r})|e^{ik|\mathbf{r}|}$ with $\mathbf{r}$ a vector point in space with for instance the center of slit $A$ as the origin. For a particular $\mathbf{r}$, $|\Psi(\mathbf{r})|^2 d\mathbf{r}$ represents the probability of finding the quantum system's particle-part in the volume element $d\mathbf{r} \equiv dxdydz$ and the phase factor $e^{ik|\mathbf{r}|}$ represents the phase of the accompanying non-empty de Broglie wave, i.e. the phase of the quantum system. This interpretation can also be associated to $\Psi_{single,A}$. However, for the interpretation of $\Psi_A$ and $\Psi_A$ the situation is different for the simple reason that $\Psi_A$ and $\Psi_A$ are functions representing *empty* de Broglie waves. As the empty de Broglie waves are not accompanied by any particle-part of a quantum system, $|\Psi(\mathbf{r})|$ has no relevance and is therefore normalized to a constant, in our case being $\alpha$ and $\beta$. In fact, with $d<<D$ we may put $\alpha=\beta=1$ all the time without loss of generality. On the other hand, the phase $e^{ik|\mathbf{r}|}$ of the empty de Broglie waves is absolutely relevant so that $\Psi_A=e^{ikr_A}$ and $\Psi_B=e^{ikr_B}$ are actually phase functions. In fact, it would be better to use a different representation, such as $\wp_A=e^{ikr_A}$ and $\wp_B=e^{ikr_B}$, just to emphasize that those are phase functions and not ordinary wavefunctions. However, for the sake of generality and the fact that the use of wavefunction symbols is most common, I consistently use the symbols $\Psi_A$ and $\Psi_B$ as phase functions.

Just for the record, from Eqs. (12) and (13) we can reconstruct easily the usually observed two-slit interference pattern in case the photons' particle-part sometimes passes slit *A* and sometimes slit *B*, i.e. the laser spot covers both slits *A* and *B* simultaneously (see Fig. 3). As the two-slit interference is in fact about self-interference, the intensity on the screen due to the photons' particle-part emanated from slit *A* can simply be added to the intensity due to the photons' particle-part emanated from slit *B*. The resulting two-slit interference in Fig. 3(b) is then given as the sum of Eqs. (12) and (13), i.e.

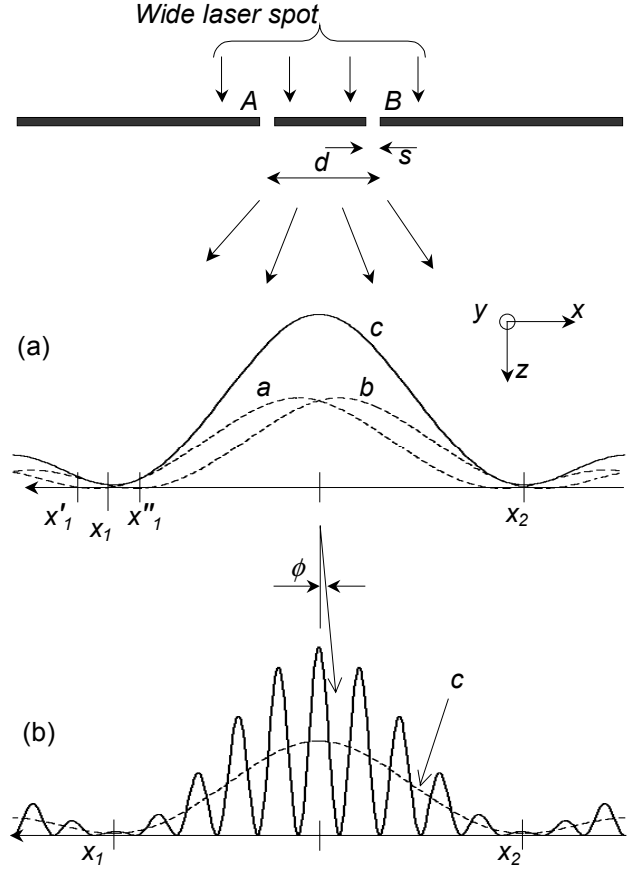

Figure 3: The usual two-slit interference experiment: (a) Curves *a* and *b* represent the single-slit diffraction intensity pattern from their respective single slit. Single-slit-like curve *c* represents the sum of curves *a* and *b* without two-slit interference; (b) The final two-slit interference pattern is given by the solid-line curve. Note that $x_1$ and $x_2$ are shifted over a distance $d/2$ compared to $x_1$ and $x_2$ in Fig. 2.

$$I_{A-B}(x) = I_0 \left\{ \left[ \frac{\sin(\pi s(x+d/2)/\lambda D)}{\pi s(x+d/2)/\lambda D} \right]^2 + \left[ \frac{\sin(\pi s(x-d/2)/\lambda D)}{\pi s(x-d/2)/\lambda D} \right]^2 \right\} \cos^2\left(\frac{\pi x d}{\lambda D}\right) \quad (14)$$

where the part $\{[\sin()/()]^2+[\sin()/()]^2\}$ is depicted by the curves *c* in Fig. 3. An interesting finding maybe is the fact that the minimum intensity of the term in $\{-\}$ is different from zero and can be calculated as follows. In Fig. 3(a), $x'_1$ is the co-ordinate of the minimum of curve *a* which can be derived from the condition that $\pi s(x'_1+d/2)/\lambda D = -\pi$ resulting in $x'_1 \approx -d/2-\lambda D/s$. Consequently, from $\pi s(x''_1-d/2)/\lambda D = -\pi$ one can derive that $x''_1 \approx +d/2-\lambda D/s$ which is the co-ordinate of the minimum of curve *b*. To a good approximation we can represent the co-ordinate of the minimum of curve *c* in Fig. 3(a) as $x_1 \approx (x'_1+x''_1)/2 \approx -\lambda D/s$. Substituting $x_1 \approx -\lambda D/s$ in the first sin()/() term of Eq. (14) results in

$$\left[\frac{\sin(-\pi+(\pi sd/2\lambda D))}{-\pi+(\pi sd/2\lambda D)}\right]^2 = \left[\frac{-\sin(\pi sd/2\lambda D)}{-\pi+(\pi sd/2\lambda D)}\right]^2 \approx \left(\frac{sd}{2\lambda D}\right)^2$$

under the condition that $\pi sd/2\lambda D \ll 1$ which is clearly the case for a two-slit photon interference experiment with realistic values for *s*, *d*, $\lambda$ and *D*. With $\lambda = 0.63\,\mu\text{m}$, $s = 2\,\mu\text{m}$, $d = 12\,\mu\text{m}$ and $D = 10\,\text{cm}$, we find that $(sd/2\lambda D)^2 I_{0,\text{single}} \approx 4\times 10^{-8} I_{0,\text{single}}$ which is the from slit *A* originating intensity at $x_1$ for curve *c*. The intensity at $x_1$ coming from slit *B* can be calculated in the same way and is less than $4\times 10^{-8} I_{0,\text{single}}$ so that the total intensity at $x_1$ for curve *c* in Fig. 3(a) is in the order of $10^{-7} I_{0,\text{single}}$. Now, this value is indeed different from zero but nonetheless so small that, in conjunction with the fact that $|x_1| \approx \lambda D/s \gg d/2$ for $D = 10\,\text{cm}$, curve *c* can be interpreted as a normal single-slit diffraction pattern with the single slit at $x = 0$ and which has a width of *s*. Hence, in a two-slit interference experiment in Fraunhofer regime, Eq. (14) can be reduced to the usually observed approximate relation

$$I_{\text{A-B}}(x) = I'_0 \left(\frac{\sin(\pi xs/\lambda D)}{\pi xs/\lambda D}\right)^2 \cos^2\left(\frac{\pi xd}{\lambda D}\right) \qquad (15)$$

with $I'_0 \approx 2I_0 \approx 4I_{0,\text{single}}$. Furthermore, also in Fig. 3(b) Heisenberg's single-slit uncertainty relation $s\Delta p \geq h$ remains upheld as most of the intensity is spread out but still mainly remains within the range $x_1$, $x_2$.

## 3 Discussion

Despite the fact that past attempts for directly detecting empty de Broglie waves have failed [36,42,43], the intensity pattern in Fig. 2(b) of the proposed experiment would, although indirect, be a clear and nonetheless strong indication of the existence of empty de Broglie waves. In recent decades empty de Broglie waves and experimental proposals and implementations for its detection have been discussed in several debates and publications [36,42–48] (see especially Chapter 3 in Lucien Hardy's Ph.D. thesis [44] and references therein). The reasoning for the case of this experiment is clear: if the particle-part of the photons always pass slit *A* while observing the two-slit interference as in Fig. 2(b), then the only conclusion is that there must have passed an empty de Broglie wave through slit *B* which then interacted somehow with the particle-part behind the two-slit plate.

Another consequence of the expected pattern in Fig. 2(b) with far-reaching implications is related to the visibility-which-way inequality relations [16–22]. There are mainly three inequalities:

$$P^2+V^2 \leq 1, \; K^2+V^2 \leq 1 \text{ and } D^2+V^2 \leq 1 \tag{16}$$

with *P* the which-way predictability, *K* the which-way knowledge, $D=max\{K\}$ the which-way distinguishability and *V* the visibility defined in the standard way as

$$V = \frac{I_{max} - I_{min}}{I_{max} + I_{min}} \tag{17}$$

with $I_{max}$ the maximum and $I_{min}$ the minimum intensity of the two-slit interference pattern.

Inequalities (16) constitute the generally accepted principle that one cannot have simultaneous absolute knowledge about which way (slit) the interfering quantum system's particle-part took in a two-slit or a Mach-Zehnder type interference experiment, while maintaining interference with maximum visibility. The quantities *K* and *D* are the result of an effectively performed direct which-way measurement on the particle-part of the quantum system so to determine which way the quantum system took [21,22]. However, if one can predict by the construction of the experimental set-up – so, without performing any direct which-way measurement on the particle-part – what way (slit) the particle-part of the quantum system took, then one speaks about the predictability *P* of the quantum system going one way or the other [19]. As it is already determined in Fig. 2 by construction in what slit the photons' particle-part will enter even before they have left the laser, it is obvious to use *P* (instead of *K* or *D*) so that for Fig. 2(b) the relation $P^2+V^2 \leq 1$ is the one to be scrutinized.

Suppose now that in Fig. 2 the laser beam is expanded and would cover both slits; then one would not know what slit the individual photons pass through. The probability that an individual photon passes through slit *A* would be equal to the probability that it passes slit *B*. One would not be able to predict what slit the individual photon took so that $P=0$. If, on the other hand, the laser beam is focused onto slit *A*, as in Fig. 2, so that the particle-part of all individual photons pass always slit *A*, then we can predict with a 100% certainty that all photons will pass slit *A*, i.e. $P=1$, without the application of any form of direct which-way measurement on the particle-part of the photons.

However, the inequality itself in the relation $P^2+V^2 \leq 1$ does not apply to the case of the two-slit experiment in Fig. 2. The reason for that is as follows. The relation $P^2+V^2 \leq 1$ has been developed and studied in the context of a Mach-Zehnder type interferometer; see Ref. 19. If in one of the MZI paths a 100% absorber is put, it is then known with a 100% certainty which path the particle took, namely the open path, upon detection behind the output beam splitter. Consequently, we have absolute knowledge about which path the particle went without any form of which-way interaction on the particle-part itself in the MZI, hence $P=1$. Nonetheless, the interference is totally lost ($V=0$) due to the fact that the empty de Broglie wave has been completely stopped by the absorber. This implies a strong intervention into the interference process. Note that, as mentioned in section 2.2, if a particle is stopped by an absorber, then so will its empty de Broglie wave be stopped the same way. So, even though for $P=1$ no direct which-way measurement was performed on the particle-part, the reduction to zero of the empty de Broglie wave's amplitude disables completely the formation of any interference at the output beam splitter; so $V=0$.

In the two-slit experiment in Fig. 2 the situation is completely different. Here also $P=1$ (always) but, most importantly, without touching or disturbing in any way the from slit $B$ emanated empty de Broglie wave so that its amplitude remains maximum ($\beta \approx \alpha$). Consequently, as there is no intervention whatsoever in the interference process, this implies that at the screen maximum interference visibility is maintained, $V=1$. Hence, for the two-slit intensity pattern in Fig. 2(b) we then find that simultaneously $P=1$ and $V=1$ resulting in the expression

$$P^2+V^2=2 \qquad (18)$$

which clearly violates inequality (16). From this we see that which-way information and two-slit interference are not complementary at all. In fact, in this which-way experiment there is no relationship anymore whatsoever between $P$ and $V$ – i.e. they are both maximum ($=1$) – so that we can state that which-way information has become fully uncoupled from two-slit interference.

Does this mean that the standard quantum theory (SQT) and the relation $P^2+V^2 \leq 1$ are wrong somehow? Definitely not! It is just that the SQT cannot predict the outcome in Fig. 2(b) for the simple reason that SQT is not developed to describe the experiment in Fig. 2(a). The SQT tacitly assumes that it is not known which slit (or which way) the quantum system took in case of 100% visibility in the interference pattern. Usually, if one wants to know or predict which way the quantum system took, then one has to intervene somehow in the interference process (for instance by adding a which-way marker, blocking an interference path or using a second particle with which the interfering particle can interact and exchange information) which in turn, as SQT predicts, destroys the interference due to, for instance, too large momentum kicks or the introduction of orthogonality between two paths. SQT *cannot* describe the situation where it is known with certainty which way the quantum system took *without* intervening in the interference process itself. So, SQT describes quantum phenomena under specific experimental conditions and the experiment in Fig. 2 goes beyond those conditions, i.e. in Fig. 2 we know what slit the quantum systems take without intervening in the interference process. It is just that SQT has it limitations and experimental confirmation of Fig. 2(b) would be a strong indication that SQT is *not* wrong, though, but simply an incomplete theory which needs to be extended in order to be able to describe the experimental set-up in Fig. 2(a) and its outcome in Fig. 2(b).

## 4 Conclusion

A novel which-way experiment has been discussed which, upon experimental confirmation of its predicted result in Fig. 2(b), may shed light from a quite different angle on the duality relation(s) $P^2+V^2 \leq 1$ and the notion of complementarity. Apart from the experiment itself, the fundamentals of a new particle-wave model have been introduced and discussed qualitatively. There is no doubt that an experimental confirmation of the predicted outcome in Fig. 2(b) would be of great value and with far-reaching implications for the interpretation of the foundations of quantum and optical theories. As Fig. 2(b) cannot be predicted by standard quantum theory nor by any classical optical nor by any quantum optical theory, confirmation of Fig. 2(b) would inevitably spark the debate about the incompleteness of those theories. The basic laser technology for the realization of one-dimensional super-focusing is definitely available. The only challenge left in the laboratory is realizing this form of (super-)focusing with the smallest focusing angle $\theta$ possible, or at least with a $\theta$ of an order of magnitude smaller than $\phi$ ($\sin\phi=\lambda/2d$) in Fig. 2(b).

**Additional information concerning references**

[8, 21, 22]    http://www.mpq.mpg.de/qdynamics/members/index.html

[12]           http://www.physics.gatech.edu/ultracool/Papers/scattering_ifm_prl95.pdf

[26, 27, 28]   http://www.optik.uni-erlangen.de/leuchs/inmik/englisch/index.html

[29]           http://www.st-andrews.ac.uk/~atomtrap/papers/AJPBessel.pdf

[30]           http://www.st-andrews.ac.uk/~atomtrap/papers/Bessel_CP.pdf

[31]           http://www.iop.org/EJ/abstract/1367-2630/6/1/136

[41]           http://www.physicsessays.com/catalog.asp?code=1604

[44]           http://catalogue.bl.uk/F/?func=file&file_name=login-bl-list ; then Search the Integrated Catalogue for 'Hardy, Lucien'.

# Referee's comments on the paper

For those comments the reader is referred to the original journal's paper: Vol. III, 2006, no3; http://merlin.fic.uni.lodz.pl/concepts/index.htm

# Reply to referee's comments by the author

## Preface

First of all I would like to thank the referee for the effort and the time she or he has put in compiling the comments on my paper. It happens all too often that referees get rid of the job all too easy by simply rejecting the paper and not justifying why exactly the paper is to be rejected. I appreciate very much the referee's effort.

## 1  Content of the comments

So, hereby I would like to clarify a few things in my paper and at the same time reply to the referee's comments. On page 2 of the comments the referee says '*The author does not discloses his "know how" for calculations of $\Psi_{single,A}$ and $\Psi_A + \Psi_B$, so I shall do it instead of him.*' I would like to say that those calculations are not at all the essence of my paper. The calculations leading to Eqs. (I6), (I8) and (I9) can be found in any textbook; see Alonso & Finn (Ref. 40 in my paper), *Principles of Optics* from M. Born & E. Wolf, and *Introduction to modern optics*, by Grant R. Fowles.

The essence is the connection between $\Psi_{single,A}$ and $\Psi_A + \Psi_B$, and that connection is a superposition; i.e. $\Psi_A + \Psi_B$ is superimposed onto $\Psi_{single,A}$ resulting in

$$\Psi_{t,A} = N\Psi_{single,A}(\Psi_A + \Psi_B) \tag{R1}$$

where $\Psi^*_{t,A}\Psi_{t,A}$ describes the interference pattern in Fig. 2(b) when the beam is focused on slit *A* and all particle-parts pass slit *A*. From Eq. (I9) and the few lines of following comment, it looks like the referee thinks that Eq. (I9) and Fig. 3(b) are the essence of my paper. That is definitely not the case. Fig. 2(b) and Eq. (1) in my paper are the essence. Fig. 3(b) is just a consequence of the interference pattern in Fig. 2(b).

In the conclusion the referee says that the experiment is not about a super-focused laser beam as $\psi_0(r')$ =constant over the two slits. That conclusion clarifies to me why the referee sees the whole issue different than I do. I would like to point out that $\Psi_A$ and $\Psi_B$ are actually not wavefunctions; it is better to call them phase functions – because that is what they are – and that is something very different than a wavefunction. The issue of phase functions has been elaborated in the text of my paper between Eqs. (13) and (14).

Apparently, the referee overlooked that and is probably the reason for the confusion. So symbols $\Psi_A$ and $\Psi_B$ are best replaced by another symbol, say $\wp_A$ and $\wp_B$, so that

$$\wp_A = \alpha\, e^{ikr_A}, \quad \wp_B = \beta\, e^{ikr_B} \tag{R2}$$

As mentioned in my paper, the phase functions represent the phase of the empty and non-empty de Broglie wave. So, Eq. (R1) can be written as

$$\Psi_{t,A} = N\Psi_{single,A}(\wp_A + \wp_B) \tag{R3}$$

Note that Eq. (R3) also holds for a grating; in that case $\wp_A + \wp_B$ is then replaced by $\wp_A + \wp_B + \wp_C + \wp_D + \ldots$ The amplitude of the de Broglie wave in any spatial vector point, at a distance **r** away from the particle-part, rolls off as **r** increases, see section 2.2 and Fig. 2(a) in my paper. The further away from the particle-part, the smaller the amplitude of the de Broglie wave is. However, the distance between the two slits (in the order of $\mu$m) is so small that when the particle-part of the quantum system passes slit $A$, then the amplitude of the accompanying empty de Broglie wave in slit $B$ will be only slightly less than the amplitude of the non-empty de Broglie wave at slit $A$. This finding is supported by the fact that two-slit self-interference with only one quantum system at the time in the interferometer results in a two-slit interference pattern with a 100% visibility. Hence, I take the amplitude of the de Broglie wave at slit $A$ and slit $B$ approximately equal, i.e. $\alpha \approx \beta$. Hence, we can shift $\alpha$ and $\beta$ into the factor $N$ in Eq. (R3) by setting $\alpha = \beta = 1$.

The fact that the de Broglie wave stretches out in space far away (in the order of many $\mu$m) from the particle-part has a consequence on the interpretation of the wavefunction. Have a look at Fig. 2(a). As the corpuscle photon particle-parts of the laser beam are focused onto slit $A$, this consequently entails that at the two-slit plate the wavefunction $\psi(\mathbf{r})$ in standard quantum theory (SQT) will be concentrated around slit $A$ while at slit $B$ the amplitude of the wavefunction will be practically zero, i.e. $\psi(A) =$ maximum and $\psi(B) \approx 0$. Although $\psi(B) \approx 0$, there is still two-slit interference for the simple reason that the accompanying de Broglie waves stretch out in space far beyond and outside $\psi(\mathbf{r})$. In Fig. 2(a) there is no relationship between the amplitude of $\psi(\mathbf{r})$ being confined around slit $A$ and the amplitude of the de Broglie wave at slit $B$. From the referee's conclusion it is clear the referee considers $\Psi_A$ and $\Psi_B$ in Eq. (R1) as wavefunctions while they are not. I hope this issue is now clarified with the re-introduction of phase functions.

Before continuing to the next section, I would like to address a few technicalities in the referee's comments. I am not all too sure if I understand the referee correctly in several technical issues in the comments. In the text between Eqs. (I4) and (I5) the expression $R_A \approx D - \xi(x+d/2)/D$ appears. But, with $\xi$ and $x$ positive, $R_A < D$ and that cannot be the case: $R_A$ is always $\geq D$. Then, right after Eq. (I6a), the referee states "*So diffraction on a single slit depends on choice of the coordinates origin. It cannot be.*" I am not sure what the referee means by that: replacing *x* by *x+d/2* or *x+L* means the single-slit pattern is simply shifted over a distance *–d/2* or *–L*; that has nothing to do with single-slit diffraction being depended on the coordinates' origin. Then there is also the integral in Eq. (I9). The boundaries in Eq. (I9) run from *–s/2* to *+s/2*, but that is not quite correct I think. Probably, the integral should be a sum of two integrals with boundaries from *–s/2–d/2* to *+s/2–d/2* for the first integral, and from *–s/2+d/2* to *+s/2+d/2* for the second integral; see the book of G.R.Fowles.

## 2  Quantum System function

At this point I would like to introduce and propose a whole new function, very different than a wavefunction. As discussed at the end of section 2.2 in my paper, on the level of physical reality interference is about self-interference and in my opinion the only correct way for describing interference is by describing it on the level of a single quantum system taking into account the particle-part and the accompanying de Broglie wave as one 3-dimensional object. That would probably require a whole new (numerical) theory, or at least a strongly revised quantum theory. In my view, what we in the first place need is a mathematical description of the entire single quantum system, i.e. a description that models the corpuscle particle-part and the accompanying de Broglie wave in one mathematical object, which then can be used to describe self-interference on the level of a single quantum system. So what we need is some kind of a deterministic space-time Single Quantum System function (abbreviated as *SQSfunction*), which I represent as

$$SQSfunction(\mathbf{r_0},t) = f\left(\Lambda(\mathbf{r_0},t), A_{dB}(\mathbf{r},t)\exp\{i(\mathbf{k}\cdot\mathbf{r}-\omega t)\}\right) \qquad (R4)$$

with $\mathbf{r_0}$, $\mathbf{r}$ and $\mathbf{k}$ being vectors and *t* the time. This *SQSfunction* is a function (*f*) of the right hand side factors and should describe the exact space-time trajectory of the corpuscle particle-part from laser to target screen over slit *A*. At a particular time the coordinate $\mathbf{r_0}$ represents the exact position in space of the corpuscle particle-part of the quantum system. The function $A_{dB}(\mathbf{r},t)$ is a 3-dimensional amplitude function that represents the amplitude of the de Broglie wave at any space-time point in the vicinity of the corpuscle particle-part; $\exp\{i(\mathbf{k}\cdot\mathbf{r}-\omega t)\}$ represent the phase of the de Broglie wave at the same space-time point and can be called the phase function. The function $\Lambda$ is a mathematical device that describes the connection and interaction between the corpuscle particle-part and the de Broglie wave. Note that $A_{dB}(\mathbf{r},t)$ is pure relativistic, i.e. if with respect to a rest frame the quantum system does not move, then $A_{dB}(\mathbf{r},t)\equiv 0$ $\forall \mathbf{r},t$ so that the position of the particle-part remains fixed in space at $\mathbf{r_0}$. So, a correct description of the single quantum system should be done starting from a pure relativistic framework.

Eq. (R4) can also be written without the time *t*. Let us have a look at Fig. 2(a) in my paper. Consider now a photon particle-part that leaves the laser, passes slit *A* and then finally hits on the target screen. Assume then for a moment that one can take a series of pictures, one after another, of the same photon particle-part once it left the laser until the particle-part hits the target screen as in Fig. 2(a). If we then put all the pictures on top of each other, we then get one continues string (line) of corpuscle particle-parts starting from the laser, through slit *A*, which then ends at a particular point at the target screen. This string of particle-parts is also joint by its accompanying de Broglie wave that stretches from the laser to slit *A*; from slit *A* to the target screen this accompanying de Broglie wave then evolves to a two-slit interference pattern that pushes and pulls the single particle-part trajectory so to hit the screen at a particular coordinate. This de Broglie wave from laser to target screen can be interpreted as a very long-stretched de Broglie wave 'field'. In fact, this whole 'spatialization' procedure of eliminating time in Eq. (R4), thereby making the expression purely spatial, is already depicted in Fig. 2(a) but with only a few particle-parts. Imagining Fig. 2(a) with a string of particle-parts starting from the laser, going over slit *A* until it reaches a particular point at the target screen is exactly a visualization of this 'spatialization' procedure. After this procedure Eq. (R4) becomes

$$SQSfunction(\mathbf{r_0}) = f\big(\Lambda(\mathbf{r_0}), A_{dB}(\mathbf{r})\exp(i\mathbf{k}.\mathbf{r})\big) \tag{R5}$$

where $\exp(i\mathbf{k}.\mathbf{r})$ is the familiar phase function in Eq. (R2). Now, this method of not using time for calculating interference works very well as time is also not used in, for instance, the Fresnel-Kirchoff approach (see Born & Wolf p.425). However, one ought to remain very cautious when interpreting this spatialization procedure for the reason that the 'string' of particle-parts and the de Broglie wave 'field' as such do not exist in physical reality when dealing with a single quantum system being studied over a larger space or longer distance. Note that in Eqs. (R4) and (R5) there is no appearance of any probability amplitude function $\psi$: that is very important.

## 3 Introduction of $\psi$

Although there is no probability amplitude function $\psi$ in Eq. (R5), in my view one can anyhow introduce $\psi$ on the ensemble level. Knowing that the laser beam is focused onto slit *A*, repeat then the spatialization procedure for a large number of photons coming from the laser. This results then, not in a string but, in a beam of particle-parts between the laser and slit *A*, which then 'flutters' open beyond slit *A* in a continuum of particle-parts to finally hit the target screen according to an interference pattern. So, based on Eq. (R5) I then define a new function, namely an Ensemble of Quantum Systems function (*EQSfunction*). When considering at slit *A*, it is clear this function can be written as

30... 

$$EQSfunction(\mathbf{r_0}) = f\left(\psi(\mathbf{r_0}), A_{dB}(\mathbf{r})\exp(i\mathbf{k}.\mathbf{r})\right) \qquad (R6)$$

where the probability amplitude function $\psi(\mathbf{r_0})$ is concentrated about slit $A$. As explained under Eq. (R3), the amplitude of the de Broglie wave $A_{dB}(\mathbf{r})$, and hence the range of impact of $A_{dB}(\mathbf{r}) \times \exp(i\mathbf{k}.\mathbf{r})$, stretches out in space far wider than $\psi(\mathbf{r_0})$; that if very noticeable when $\psi(\mathbf{r_0})$ is strongly confined, as is the case in my two-slit experiment where $\psi(\mathbf{r_0})$ is confined about slit $A$; i.e. $\psi(\mathbf{r_0})_A$ = maximum and $\psi(\mathbf{r_0})_B \approx 0$, as is presented in Fig. R1(a).

Note that $\psi(\mathbf{r_0})$ in Eq. (R6) does not carry any phase information because that is already present in $\exp(i\mathbf{k}.\mathbf{r})$. That is why I call $\psi(\mathbf{r_0})$ in Eq. (R6) a probability amplitude function and not a wavefunction as in standard quantum theory (SQT). However, I see there is a relation between $\psi(\mathbf{r_0})$ in Eq. (R6) and the wavefunction $\psi(\mathbf{r_0})$ in SQT; i.e. $\psi(\mathbf{r_0})_{R6} = |\psi(\mathbf{r_0})|_{SQT}$. Noteworthy to mention is that in Eq. (R6) the variable $\mathbf{r_0}$ in $\psi(\mathbf{r_0})$ is definitely not the same as $\mathbf{r}$ in $\exp(i\mathbf{k}.\mathbf{r})$. In SQT, on the other hand, the two space variables are the same and so we have $\psi(\mathbf{r_0})_{SQT} = |\psi(\mathbf{r_0})|_{SQT} \times \exp(i\mathbf{k}.\mathbf{r_0})$; i.e. the range of impact of $\exp(i\mathbf{k}.\mathbf{r_0})$ is determined by the amplitude $|\psi(\mathbf{r_0})|_{SQT}$. In Eq. (R6), on the other hand, the range of impact of $\exp(i\mathbf{k}.\mathbf{r})$ is determined by $A_{dB}(\mathbf{r})$ and that stretches out over a larger space than $\psi(\mathbf{r_0})$; see Fig. R1(a).

## 4 The step to the "invented" expression (R3)

As mentioned earlier, I am convinced that the only correct way of calculating the interference pattern in Fig. 2(b) is by calculating it on the level of every single quantum system that comes into slit $A$, using Eq. (R4) or a similar expression. However, this probably requires a whole new (numerical) theory (not to be confused with the de Broglie-Bohm theory) which forced me (as worked out in section 2.3 in my paper) to find an alternative way for calculating the interference pattern in Fig. 2(b). The basis that I use for my alternative method is Eq. (R6). Back at Fig. 2(a), once the particle-part of the quantum system passes slit $A$, the two slits ($A$ and $B$) split the de Broglie wave in two. In Eq. (R6) this is represented by $A_{dB}(\mathbf{r})$ being split in two with approximately equal amplitude at slits $A$ and $B$. Consequently $\exp(i\mathbf{k}.\mathbf{r})$ in Eq. (R6) is then also being split in two which looked at from the target screen becomes $\exp(ikr_A)+\exp(ikr_B)$, as in Eq. (R3).

On the other hand, right after the corpuscle particle-part passed slit $A$, it will at first undergo a single slit diffraction determined by that part of the de Broglie wave that went through slit $A$. Mathematically this can be modeled by masking the de Broglie wave, i.e. taking $A_{dB}(\mathbf{r})=1$ at slit $A$ and $A_{dB}(\mathbf{r})=0$ elsewhere (slit $B$). But, as $\psi(\mathbf{r_0})$ in Eq. (R6) is maximum at slit $A$ and zero at slit $B$, this masking is automatically done by $\psi(\mathbf{r_0})_{A,B}$. The effect of this masking procedure is a pure single-slit interference pattern at the target screen. And, as the particle-part experiences first a single-slit diffraction right after slit $A$ and only further away from slit $A$ a two-slit interference, it is obvious to modulate the two-slit interference onto the single-slit diffraction. Mathematically this is accomplished by multiplying the two-slit interference term $\exp(ikr_A)+\exp(ikr_B)$ with the single-slit term $\Psi_{single,A}$, which then results in Eq. (R3). One should not forget that, after all, any two-slit experiment, where the laser beam covers both slits as in Fig. 3, results in a single-slit(-like) pattern (coming from two single-slits) modulated by two-slit interference. Hence, the conclusion that a pure single-slit pattern coming from only one single-slit can be modulated by a two-slit interference pattern should not be so hard to accept.

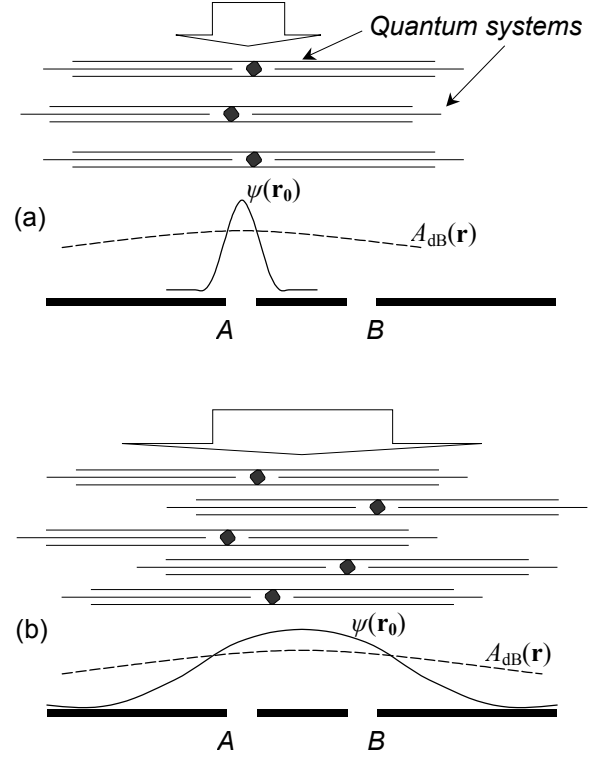

Figure R1: (a) $\psi(\mathbf{r}_0)$ covers only slit $A$ while the de Broglie wavefront with amplitude $A_{dB}(\mathbf{r})$ covers both slits. (b) $\psi(\mathbf{r}_0)$ covers both slits and thereby masks the de Broglie wavefront entirely: hence the confusion between $\psi(\mathbf{r})_{R8}=|\psi(\mathbf{r})|_{SQT}$ and $A_{dB}(\mathbf{r})$. Note that this figure is not present in the journal's version of this paper.

## 5 The confusion between $|\psi(\mathbf{r})|_{SQT}$ and $A_{dB}(\mathbf{r})$

Clearly the interference pattern in Fig. 2(b) cannot be predicted by standard quantum theory (SQT). But why does SQT, or any other wave theory for that matter, works so well for predicting two-slit interference where the laser beam covers both slits, as in Fig. 3? That can be explained using Eq. (R6). In case of a wide laser beam $\psi(\mathbf{r}_0)$ covers both slits just as the amplitude function $A_{dB}(\mathbf{r})$; see Fig. R1(b). To illustrate this better, consider the spatialization procedure that I used to come to Eq. (R6), but now on a wide laser beam of photons covering both slits *A* and *B*. This results in a coherent continuum of particle-parts that is homogenously distributed over both slits and 'flutters' then open beyond the slits towards a two-slit interference pattern on the target screen. So, beyond the slits there is a continuum of a two-slit interference de Broglie wave (field) covered by a continuum of particle-parts. Now, as the particle-parts in slits *A* and *B* are homogenously distributed, one can consider the probability amplitude $\psi(\mathbf{r_0})$=constant in both slits, just as $A_{dB}(\mathbf{r})$=cte in both slits. So, $\psi(\mathbf{r_0})$ is being masked by $A_{dB}(\mathbf{r})$ at both slits. Hence, at both slits Eq. (R6) transforms to

$$EQSfunction(\mathbf{r}) = NA_{dB}(\mathbf{r})\exp(i\mathbf{k}.\mathbf{r}) \qquad (R7)$$

where $\psi(\mathbf{r_0})$=cte has been put in the normalization constant *N*. The variable $\mathbf{r_0}$ in $EQSfunction(\mathbf{r_0})$ in Eq. (R7) has been replaced by $\mathbf{r}$ for the simple reason that $\psi(\mathbf{r_0}) = \psi(\mathbf{r})$=cte at both slits.

However, from the symbolic point of view we can, in Eq. (R7), as well replace $A_{dB}(\mathbf{r})$ by $\psi(\mathbf{r})$, which results in

$$EQSfunction(\mathbf{r}) = N\psi(\mathbf{r})\exp(i\mathbf{k}.\mathbf{r}) \qquad (R8)$$

Note that any constant change in Eq. (R8), due to the replacement of $A_{dB}(\mathbf{r})$ by $\psi(\mathbf{r})$, can be absorbed in *N*. As already pointed out right under Eq. (R6), $\psi(\mathbf{r})_{R8}=|\psi(\mathbf{r})|_{SQT}$ so that Eq. (R8) becomes

$$EQSfunction(\mathbf{r}) = N\,|\psi(\mathbf{r})|_{SQT}\,\exp(i\mathbf{k}.\mathbf{r})\,. \qquad (R9)$$

Apart from a constant, the right hand side is simply the SQT wavefunction, $\psi(\mathbf{r})_{SQT}=|\psi(\mathbf{r})|_{SQT}\times\exp(i\mathbf{k}.\mathbf{r})$. The conclusion is that, if in SQT the wavefunction is wide enough so that it covers both slits (i.e. one does not know where exactly the corpuscle particle-part is in slit *A* or *B*), then $|\psi(\mathbf{r})|_{SQT}$ literally masks completely the amplitude function $A_{dB}(\mathbf{r})$, as is presented in Fig. R1(b), so that the confusion between $|\psi(\mathbf{r})|_{SQT}$ and $A_{dB}(\mathbf{r})$ never comes at the surface and therefore never gets noticed: i.e. $|\psi(\mathbf{r})|_{SQT}$ has taken over the role of $A_{dB}(\mathbf{r})$. It is exactly this confusion that the experiment in Fig. 2(a) can bring to the surface.

Hence, the problem with SQT, and any other wave approach (like Fresnel-Kirchoff), is that those theories cannot make a distinction between $\psi(\mathbf{r})_{R8}$ and $A_{dB}(\mathbf{r})$, this becomes apparent when $\psi(\mathbf{r})_{R8}=|\psi(\mathbf{r})|_{SQT}$ does not coincide with $A_{dB}(\mathbf{r})$, as is presented in Fig. R1(a). Note that starting from a single quantum system picture in Eq. (R4) one evolves, over Eq. (R7), to a complete 'wave' (or wave 'field') picture in Eq. (R9) where the continuum of particle-parts and wave is masking the existence of any individual quantum system.

# 6 Alternative experiments

Very briefly I would like to introduce a variant of the experiment in Fig. 2, called the first alternative experiment. A possibly not so easy thing to realize might be the very small focusing angle $\theta$ in Fig. 1 in conjunction with a small spot size, as discussed in section 2.1 in my paper. There is, however, another method of realizing the experiment without having to focus the laser beam with a small focusing angle $\theta$ and accomplish the experiment with a strongly focused laser beam but where a lot larger $\theta$ is allowed. The alternative experiment I propose is the following. One can perform a wide beam experiment as in Fig. 3, not with photons although, but with atoms or electrons. Once the atoms or electrons passed slits *A* and *B*, then those particle-parts of the quantum systems that passed slit *B* can be knocked out of the interferometer in the *y*-direction in Fig. 1 using a strongly focused laser spot where the average wavevector **k** in the spot is aligned along the *y*-direction and positioned right behind slit *B*. That way, one eliminates all quantum system at the target screen that came through slit *B*. As we have not intervened into the interference process for the quantum systems passing through slit *A*, the two-slit interference pattern at the target screen is left intact with a 100% visibility (but with 50% reduced intensity) so that $P^2+V^2=2$ (or at least $P^2+V^2>1$) for we know that all detected quantum systems came through slit *A*. Note not to confuse this experiment with the proposal of Richard Feynman; see Ref. 2 in my paper.

     A second alternative experiment is based on a Mach-Zehnder Interferomenter (MZI) using atoms; see Ref. 12 in my paper. One can then knock out all atoms in one of the two MZI arms by directing an ordinary laser beam (without focusing) onto one of the MZI arms, and with the extra condition that the wavevector **k** of the laser is best perpendicularly oriented onto to the plane formed by the two MZI arms. That way the knocked-out atoms receive an additional momentum kick perpendicular to that MZI plane and thereby leave the MZI. As the laser, upon interaction with the knocked-out atoms, has interacted only with the atoms itself and not with the empty de Broglie waves in the same arm from the atoms that went the other arm, maximum interference visibility will be sustained at the MZI output (but with a 50% reduced hit-rate) so that $P^2+V^2=2$ for we know that all detected atoms went the other arm. Note that a MZI experiment like this will only work if the interference is clearly self-interference, i.e. only one atom at the time is in the MZI and the absolute difference in length between the arms must be in the order of only a few atom wavelengths. If those conditions are not met then this MZI experiment will not work.

# 7 Some general findings and conclusions

A problem in the discussion of interference is that the contemporary perception about two-slit interference on the level of physical reality is a perception solely based on a wave picture as interference is modeled that way on theoretical level; see text following Eq. (R9). The reviewer referred to calculating the two-slit interference using the Fresnel-Kirchoff (see Born & Wolf p.425) approach but this entails (only) a wave picture. So, the Fresnel-Kirchoff approach tacitly assumes that physical reality is about (only) waves. I am afraid this is only half of the truth about physical reality. The other half of truth is simply not considered in the Fresnel-Kirchoff approach and that is the particle-part. Just try to explain atom two-slit self-interference, see section 2.2 in my paper and Carnal & Mlynek (Ref. 35), on the level of a single atom and the problem becomes clear.

The Fresnel-Kirchoff approach is a fine instrument, which can only be verified IF a large number of quantum systems (quasi continuum of particle-parts) is observed at the screen and where the quantum systems have (in normal circumstances) equal probability of passing any of the two slits. The Fresnel-Kirchoff approach simply fails if we try to describe two-slit self-interference on the level of only one single quantum system that passes through the two-slit plate. The reason is that the Fresnel-Kirchoff approach cannot separate the narrow $\psi(\mathbf{r})$ around the particle-part from the wider spaced $A_{dB}(\mathbf{r})$ as it is only a wave approach.

A remark maybe is that in contemporary theories (quantum and optical) the atom is interpreted sometimes as a particle (in case of direct detection) or sometimes as a (de Broglie) wave (as in the case of interference). However, there is an angle to this generally accepted particle-wave model. From interference experiments (using atoms for instance) performed the last several decades it is clear that in physical reality the interfering atom quantum system is composed of the atom particle-part which is at all times simultaneously joint by the de Broglie wave. As just mentioned, the contemporary theories sometimes treat the atom quantum system as a particle OR sometimes as a wave (as in interference) but never as a particle AND a wave at the same time. Nonetheless, an atom quantum system is in physical reality a particle AND a wave at the same time. Quantum systems do not change their dress from particle to wave and vice-versa depending on the experimental set-up they are in or the type of theory we use.

I would like to make a last note on the issue of incompleteness of physical theories. All physical theories (classical, quantum, optical, mechanical, relativity, ..) are incomplete by definition; there is no theory with which one can describe ALL physics. With general relativity one can quantify the strength of, for instance, the gravitational lens effect from a large planet when that planet passes between its orbiting sun and Earth; standard quantum theory (SQT) won't be of much help in quantifying gravitational lenses. On the other hand, quantum phenomena in semiconductors for instance can be quantified by SQT; here the general relativity theory won't do you any good. So, all physical theories are incomplete and that is the most obvious and natural thing there is. Hence, we can call this: incompleteness in the obvious sense. Note that incomplete theories are NOT necessarily wrong or erroneous. It is just that one has to use the theories in the correct context, meaning the context for which the theories have been designed or developed for. There is, however, another form of incompleteness which might be more serious and which I call incompleteness in the non-obvious sense. If one can come up with a typical quantum experiment, as the one in my paper, from which the experimental outcome cannot be predicted by SQT, then this means that SQT is incomplete in the non-obvious sense. I call this form of incompleteness non-obvious for the simple reason that the experiment situates itself clearly in the quantum domain and should therefore be obviously covered by SQT. However, as it is not obvious that SQT cannot predict the results of my experiment in Fig. 2(b), this entails that Fig. 2(b) unveils a non-obvious incompleteness in SQT. This, in fact, can be considered as a (serious) shortcoming in SQT. But then again, this does not mean that SQT is erroneous – it simply means that the contemporary SQT is not designed and developed for predicting Fig. 2(b).

This concludes my reply to the referee's comments

Johan Wulleman

___